%% file: main.tex
\documentclass[conference]{IEEEtran}
\IEEEoverridecommandlockouts
\usepackage{cite}
\usepackage{amsmath,amssymb,amsfonts}
\usepackage{algorithmic}
\usepackage{graphicx}
\usepackage{graphics}
\usepackage{textcomp}
\usepackage{soul}
\usepackage{color}
\usepackage{float}

\usepackage{amssymb}
\usepackage{multirow}
\usepackage{enumerate}
\usepackage{epsfig}
\usepackage{subcaption}
\usepackage{etoolbox}
\usepackage[]{algorithm2e}
\usepackage{subcaption}
\usepackage{bbm}
\usepackage{placeins}
\usepackage{gensymb}
\usepackage{amssymb}
\usepackage{placeins}
\usepackage{caption}
\usepackage{titlesec}
\usepackage{framed,enumitem} 
\usepackage[utf8]{inputenc}
\usepackage{amsmath}

\usepackage{listings}
\usepackage{xcolor, soul}
\usepackage{tikz}
\usetikzlibrary{positioning,shapes,arrows, fit, calc}

\definecolor{codegreen}{rgb}{0,0.6,0}
\definecolor{codegray}{rgb}{0.5,0.5,0.5}
\definecolor{codepurple}{rgb}{0.58,0,0.82}
\definecolor{backcolour}{rgb}{0.95,0.95,0.92}

\lstdefinestyle{mystyle}{
    backgroundcolor=\color{backcolour},   
    commentstyle=\color{codegreen},
    keywordstyle=\color{magenta},
    numberstyle=\tiny\color{codegray},
    stringstyle=\color{codepurple},
    basicstyle=\ttfamily\footnotesize,
    breakatwhitespace=false,         
    breaklines=true,                 
    captionpos=b,                    
    keepspaces=true,                 
    numbers=left,                    
    numbersep=5pt,                  
    showspaces=false,                
    showstringspaces=false,
    showtabs=false,                  
    tabsize=2
}

\usepackage{todonotes}
\makeindex

\newcommand{\beq}{\begin{equation}}
\newcommand{\eeq}{\end{equation}}

\usepackage{url}

\def\BibTeX{{\rm B\kern-.05em{\sc i\kern-.025em b}\kern-.08em
    T\kern-.1667em\lower.7ex\hbox{E}\kern-.125emX}}

\begin{document}
\emergencystretch 3em

\title{HAPSSA: Holistic Approach to PDF malware detection using Signal and Statistical Analysis\\
\thanks{\text{Acknowledgement:} This work has been supported by the ONR contract \#N68335-17-C-0048.
The views expressed in this paper are the opinions of the authors and do not represent official positions of the Department of the Navy.
*Lakshmanan Nataraj contributed to this work while he was employed at Mayachitra.}
}

\makeatletter
\newcommand{\linebreakand}{%
  \end{@IEEEauthorhalign}
  \hfill\mbox{}\par
  \mbox{}\hfill\begin{@IEEEauthorhalign}
}
\makeatother


\author{
    \IEEEauthorblockN{Tajuddin Manhar Mohammed}
    \IEEEauthorblockA{\textit{Mayachitra, Inc.} \\
        Santa Barbara, California \\
        mohammed@mayachitra.com}
    \and
    \IEEEauthorblockN{Lakshmanan Nataraj*}
    \IEEEauthorblockA{\textit{Mayachitra, Inc.} \\
        Santa Barbara, California \\
        lakshmanan\_nataraj@ece.ucsb.edu}
    \and
    \IEEEauthorblockN{Satish Chikkagoudar}
    \IEEEauthorblockA{\textit{U.S. Naval Research Laboratory} \\
        Washington, D.C. \\
        satish.chikkagoudar@nrl.navy.mil}
    \linebreakand 
    \IEEEauthorblockN{Shivkumar Chandrasekaran}
    \IEEEauthorblockA{\textit{Mayachitra, Inc.} \\
        \textit{ECE Department, UC Santa Barbara}\\
        Santa Barbara, California \\
        shiv@ucsb.edu}
    \and
    \IEEEauthorblockN{B.S. Manjunath}
    \IEEEauthorblockA{\textit{Mayachitra, Inc.} \\
        \textit{ECE Department, UC Santa Barbara}\\
        Santa Barbara, California \\
        manj@ucsb.edu}
}

\maketitle

\begin{abstract}
Malicious PDF documents present a serious threat to various security organizations that require modern threat intelligence platforms to effectively analyze and characterize the identity and behavior of PDF malware.
State-of-the-art approaches use machine learning (ML) to learn features that characterize PDF malware. However, ML models are often susceptible to evasion attacks, in which an adversary obfuscates the malware code to avoid being detected by an Antivirus. In this paper, we derive a simple yet effective holistic approach to PDF malware detection that leverages signal and statistical analysis of malware binaries. This includes combining orthogonal feature space models from various static and dynamic malware detection methods to enable generalized robustness when faced with code obfuscations. 
Using a dataset of nearly 30,000 PDF files containing both malware and benign samples, we show that our holistic approach maintains a high detection rate (99.92\%) of PDF malware and even detects new malicious files created by simple methods that remove the obfuscation conducted by malware authors to hide their malware, which are undetected by most antiviruses.
\end{abstract}



\section{Introduction}
The widespread adoption of Portable Document Format, commonly known as PDF, is due to its portability and its intrinsic flexibility.
In fact, the PDF files can contain a variety of media including text, images, and also embedded malicious Shockwave Flash (SWF) files and ActionScript codes that will be executed by the PDF reading software.
The malicious actors have exploited this presence of vulnerabilities in mainstream PDF readers, to make PDF become an extremely successful vector for malware diffusion. 
Today, documents are one of the most common ways malware is spread across the Internet.
Products like Adobe PDF added macro and scripting capabilities making it possible for documents to work in much the same way as executable programs, right down to the ability to run processes and install malicious code on user systems.
Different antivirus products detect many different attack vectors, and stop infected documents before they reach end-user systems.
And while antivirus vendors continually try to patch the holes malware writers use to spread their code, they are usually well behind the bad guys.

To address this issue, the malware research community produced several solutions for detecting malicious PDFs in the recent past. The most recent and promising solutions~\cite{Singh2020MalwareDI,Elingiusti2018PDFMalwareDA} use various  techniques borrowed by standard malware analysis best practices like static and dynamic code analysis, and adapted to the specificities of the PDF file format to extract analyzable features for detecting malware. 
Some of widely used features include the PDF file structure information, metadata information, encoding method, and lexicon-based features. Although such hand-crafted features have shown successful results, it generally requires time and effort to design these features.
The proposed holistic approach uses the input data characteristics (i.e., byte sequences) and the file structure of PDF to effectively detect PDF malware. 
The developed features are at best \emph{orthogonal}~\cite{mohammedmalware} and can uniquely represent the PDF file in its entirety for the purpose of malware detection.
As we target the PDFs in this work, we review the structure of the PDFs and illustrate how we design a bag-of-words~\cite{Peng2016BagOV} model to look at potential risky PDF code elements.
We also develop a similar bag-of-words model by considering the malicious API call sequences in the dynamic analysis reports.
We generate obfuscation-removed samples as a way to not only augment our PDF malware dataset for the task of malware detection, but also to perform adversarial analysis determining robustness to small PDF code changes that is shown to fool other antiviruses.

In the following section, we present related work on PDF malware detection followed by description and evaluation of our proposed holistic approach to PDF malware detection.
The main contributions in this paper are as follows:
\begin{itemize}[noitemsep,topsep=0pt]
    \item We determined the best set of signal and statistical features that enable robust PDF malware detection.
    \item We proposed a holistic approach that combines the orthogonal feature space models for improved detection performance.
    \item We identified the key structural elements in both the PDF code and the dynamic analysis reports, and developed powerful bag-of-words classification models.
    \item We developed strategies that remove the obfuscation conducted by malware authors to not only perform adversarial analysis but also suggest it as a way to augment the PDF malware datasets. 
\end{itemize}




\section{Related Work}
\label{s:rw}
Typical features extracted from malware binaries can be broadly grouped into either static or dynamic features.
As the name suggests, static features are extracted from the malware without executing it. 
Dynamic features, on the other hand, are extracted by executing the code, usually in a virtual environment, and then studying their behavioral characteristics such as system calls trace or network behavior. 

Static analysis can further be classified into static code based analysis and non-code based analysis.  
Static code based analysis techniques study the functioning of an executable by disassembling the executable and then extracting features. 
A common static code based analysis is control flow graph analysis~\cite{ding2014control,ma2019combination}.
After dis-assembly, the control flow of the malware is obtained from the sequence of instructions and graphs are constructed to uniquely characterize the malware.
Static non code-based techniques are based on a variety of techniques: n-grams~\cite{liangboonprakong2013classification},  n-perms~\cite{karim2005malware}, hash based techniques~\cite{arefkhani2015malware,namanya2020similarity}, PE file structure~\cite{belaoued2016chi,rezaei2020efficient}  or signal similarity based techniques~\cite{malware-images,Nataraj2016SPAMSP}.
Features are then extracted from these to characterize the malware.
Among hash based methods, ssdeep~\cite{kornblum2006identifying} is a common technique to compute context triggered piece-wise hashes on raw binaries.
Pehash~\cite{wicherski2009pehash}, however, uses the Portable Executable (PE) file structure to compute a similarity hash.
Image similarity based methods~\cite{malware-images,mohammedmalware} convert a malware binary to a digital image and apply image processing based techniques to compute similarity features.
Audio similarity based methods~\cite{nataraj2020malware,azab2020msic} that represents malware bytes as an audio signal were also developed recently for malware detection.
Among dynamic analysis, the most common method is to execute the malware in a controlled environment and then study its execution behavior. 
Behavioral profiles or graphs are generated to build models of malware~\cite{Kolbitsch,park10}.
Some works generate a human readable report of the execution flow and extract features from the reports~\cite{Rieck08,malheur}.

Recently, there has been a considerable work involving malware analysis in PDF files~\cite{Singh2020MalwareDI,Elingiusti2018PDFMalwareDA,Jeong2019MalwareDO}.
Signature-based detection used to be the standard in cyber-security, where researchers used signatures in PDF files to identify malware~\cite{Rautiainen2009ALA}.
Many static and dynamic analysis methods~\cite{Smutz2012MaliciousPD,Srndic2013DetectionOM,Tzermias2011CombiningSA,Maiorca2012APR} for PDF malware detection focus their feature extraction on metadata and hierarchical structure of PDF document. 
Recently, a image visualization based method~\cite{Corum2019RobustPM} was also shown as a viable solution for PDF malware detection.
We emphasize that all these approaches require prior domain knowledge as they are based either
on manual feature selection or at least on the parsing of the PDF structure.
This avoids using data-hungry models such as deep neural networks (DNNs) to learn features on its own for PDF malware detection.
In this paper, we demonstrate a simple yet novel holistic approach to effectively combine these various \emph{orthogonal} features to detect PDF malware enabling generalized robustness when faced with code obfuscations. 
An orthogonality measure in the form of a joint feature score (JFS) metric~\cite{mohammedmalware} is used to explain combining these various features.
The proposed approach also has all the positives of static analysis and dynamic analysis methods while addressing some of the limitations of the aforementioned works like high time complexity, low scalability and high total feature selection count.

\section{Detection of Malicious PDF Files}
\label{s:pdf-det}

\begin{figure*}[!htbp]
    \centering
    \includegraphics[width=1\textwidth]{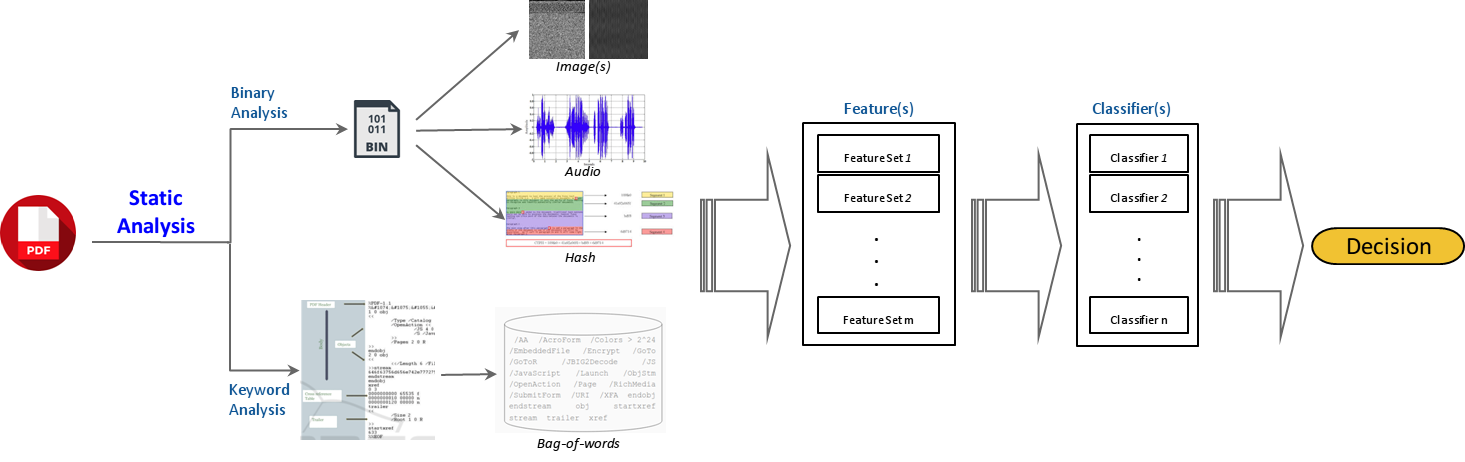}
    \caption{A holistic static analysis-based approach to PDF malware detection using binary and keyword analysis.}
    \label{sfig:comp1}
\vspace*{-0.2in}
\end{figure*}

The above approaches used in the state-of-the-art to identify PDF malware vary widely in their details and approach.
However, most of these methods incorporate a feature extraction phase in which either the PDF characteristics and/or the entire file byte-streams are analyzed for signature extraction.
To develop our holistic approach for PDF malware detection, we
consider a  categorization of the existing works with respect to (a) features used, and (b) methods to analyze the computed features.
These two aspects provide complementary information about how the existing solutions tackle the problem of identifying malicious PDF files.

As shown in Figure~\ref{sfig:comp1}, the proposed pipeline includes (1) a binary analysis component that looks at the PDF binary file as a whole to extract relevant features, and (2) a keyword analysis component that looks at the PDF file code structure to identify malicious signatures.
We then exploit the orthogonality among the features and combine them to derive a holistic model for PDF malware detection. 
An alternative approach is also proposed that includes analyzing the dynamic analysis reports of PDF files for potential malicious signatures. 
Finally, a obfuscation removal strategy is studied to disable execution of malicious code during runtime and to also perform adversarial analysis.

\subsection*{Dataset}
\label{ss:data}
To evaluate our malware detection models, we use an internal PDF malware dataset curated from various sources including samples from Contagio~\cite{contagio}.
The benign and malware labels  were validated using VirusTotal~\cite{virustotal}, Cuckoo Sandbox~\cite{cuckoo}, and several other antiviruses.
Our final dataset contains 30,584 PDF files with 12,519 samples labeled as malware and the rest of the 18,065 samples labeled as benign.

\subsection{Binary Analysis}
\label{ss:pdf-exp1}

To analyze the PDF file as a whole, we make use of the image, audio and hash representations of binaries.
For image features, we consider byteplot~\cite{malware-images,Corum2019RobustPM} and bigram-dct~\cite{mohammedmalware} image binary representations and extract the 320-dimensional global image GIST~\cite{oliva2001modeling} descriptors for each representation.
For audio features, we represent the PDF binary as a audio signal~\cite{nataraj2020malware} and extract 12-dimensional Chroma~\cite{muller2005audio}, 20-dimensional Mel Frequency Cepstral  Coefficients (MFCC)~\cite{tiwari2010mfcc}, and 128-dimensional Melspectrogram~\cite{shen2018natural} features.  
Finally, for the hash features, we consider the ASCII values of the characters of Ssdeep fuzzy hash~\cite{kornblum2006identifying} and then extract 40-dimensional feature vectors.
The number of dimensions for each of these features were empirically determined and the default parameters were generally considered in the existing Python implementations of these feature extraction methods.

\subsection{Keyword Analysis}
\label{ss:pdf-exp2}
The PDF code typically contains four sections - header, body, cross-reference table and trailer.
Generally, the body of the PDF code contains the actual malicious content. A PDF reader uses the keywords embedded in the document to understand what actions to execute.
Therefore, the set of keywords (tags) embedded in a PDF file can be an effective indicator of its behaviour. 
Inspired from the works~\cite{Pareek2013MaliciousPD,Laskov2011StaticDO,Srndic2013DetectionOM} that identify the most characteristic keywords by examining its occurrences in malicious and benign PDFs, we look at some of these tags.
Almost every PDF documents will contain the following 7 tags in the code: $obj, endobj, stream, endstream, xref, trailer, startxref$. And typically, the risky PDF object tags which holds the malicious content include:

\begin{itemize}[noitemsep,topsep=0pt]
  \item $/Encrypt$ -- indicates that the PDF document has DRM or needs apassword to be read. 
  \item $/AA, /OpenAction$ -- specifies script or action to run automatically.
  \item $/JS, /JavaScript$ -- indicates that the PDF contains JavaScript.
  \item $/RichMedia$ -- is for embedded Flash.
  \item $/Launch$ -- can launch a program or open another document.
  \item $/URI$ -- can access a resource by its URL.
  \item $/SubmitForm, /GoToR$ -- can send data to URL.
  \item $/JBIG2Decode$ -- indicates if the PDF document uses JBIG2 compression.
  \item $/XFA$ -- is for XML Forms Architecture.
  \item $/ObjStm$ -- can hide objects inside object stream.
\end{itemize} 

Based on the above information, we designed a bag-of-words model~\cite{Peng2016BagOV} that extracts a 25-dimensional feature vector obtained by the instruction counts of the risky object tags. These tags include: 
\textit{/AA, /AcroForm, /Colors, /EmbeddedFile, /Encrypt, /GoTo, /GoToR, /JBIG2Decode, /JS, /JavaScript, /Launch, /ObjStm, /OpenAction, /Page, /RichMedia, /SubmitForm, /URI, /XFA, endobj, endstream, obj, startxref, stream, trailer, xref}.

\subsection{Holistic Feature Fusion}
\label{sec:pdf-combi}

As shown in Figure~\ref{sfig:comp1}, the inferences from different malware detection methods can be combined to derive a integrated confidence score for malware detection. 
The motivation comes from ML model performance-boosting methods ~\cite{mohammedmalware,mohammed2018boosting} where \emph{orthogonality} between the features is exploited to learn complementary information provided by different feature space models.
The joint feature score (JFS) metric~\cite{mohammedmalware} is used to determine which set of features enable feature space model fusion to develop a more complete, integrated model for PDF malware detection. 
The proposed holistic approach uses the most simple feature fusion strategy in the form of feature concatenation, as will be discussed in Section~\ref{s:pdf-exp}.

\subsection{Dynamic Analysis}
\label{sec:dynamic}

\begin{figure}[!htbp]
\centering
\includegraphics[width=1\columnwidth]{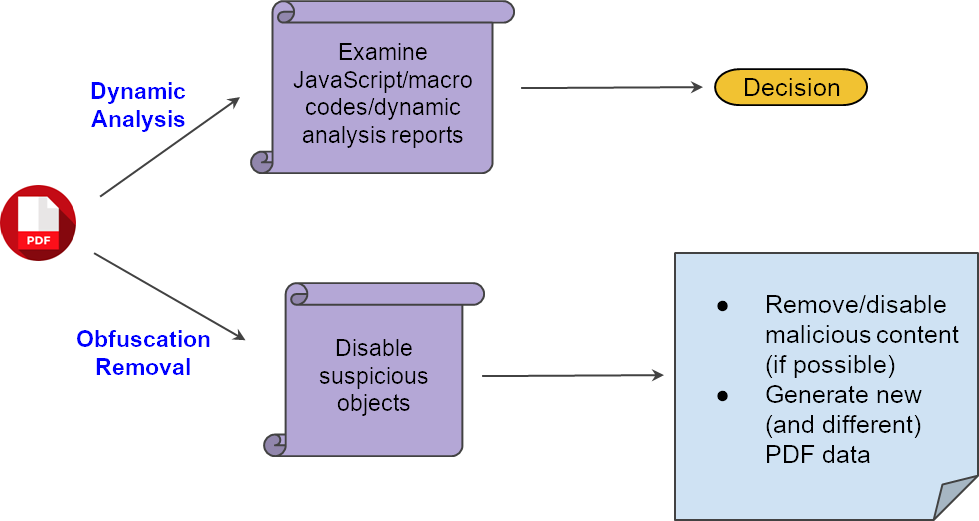}
\caption{A generic dynamic analysis-based method for PDF malware detection and description of a obfuscation removal strategy that primarily disables the execution of malicious code during runtime.}
\label{sfig:comp2}
\vspace*{-0.2in}
\end{figure}

Dynamic analysis involves running the malware sample and observing its behavior on the system in order to remove the infection or stop it from spreading into other systems.
Unlike static analysis, it's behavior-based so it's hard to miss important behaviors. 
For a comparative analysis and to test the developed principles above, we use a dynamic analysis-based method using a bag-of-words model similar to the one defined in Section~\ref{ss:pdf-exp2} to derive useful signatures for malware detection.

A generic dynamic analysis-based PDF malware detection method involves either examining embedded javascripts and macro codes in PDF files, or executing the file in a sandboxed environment for logging malicious signatures in a report, as shown in Figure~\ref{sfig:comp2}.
The proposed dynamic method derives a bag-of-words model based on the status of API call sequences in the Cuckoo dynamic reports. 
For generating the dynamic reports, we use a modified version of the Cuckoo Sandbox~\cite{cuckoo2} that has a number of advantages over the upstream Cuckoo~\cite{cuckoo}, and look at the status of various API call sequences.
This includes 184 API calls with a status of ``1'' and 73 API calls with a status of ``0''.
Overall, we form a 257-dimensional feature from the status counts of these various API calls for each sample in our PDF malware dataset.

\subsection{Obfuscation Removal Strategy}
\label{sec:pdf-deob}

As PDF language is case sensitive, any change in the object tags in PDF code will disable the corresponding objects during execution time. This is one of the most simple, yet effective obfuscation removal techniques to disable malicious content from being executed in PDF files. 
As shown in Figure~\ref{sfig:comp2}, the goal here is to determine if such a technique would indeed ``clean" the malware PDF file, and if it can be used to generate more data (augmentation) for training data-driven ML models.

In our PDF malware dataset, most of the malicious content was injected as embedded javascripts and shell scripts. 
We used two different methods to disable such objects in the code, making use of the case sensitive characteristics of PDF code.
\begin{itemize}[noitemsep,topsep=0pt]
    \item \textit{Method 1:} 
    Here, we just invert character cases of the object tags to effectively disable the object during execution. For example, $/AA$ is changed to $/aa$, $/JavaScript$ is changed to $/jAVAsCRIPT$, and so on. The objects that were disabled include: $/AA$, $/OpenAction$, $/JS$, $/JavaScript$, $/RichMedia$, $/Launch$, and $/JBIG2Decode$. 
    
    \item \textit{Method 2:}
    This is similar to Method 1, but the only difference here is that $\_disarmed$ is added as a suffix to the inverted character object tags. For example, $/AA$ is changed to $/aa\_disarmed$, $/JavaScript$ is changed to $/jAVAsCRIPT\_disarmed$, and so on.
\end{itemize}
Using the 2 strategies, we were able to generate nearly 28,000 new samples with distinct file hashes.
Therefore, this can also be regarded as an augmentation method that malware researchers can use to augment their PDF datasets.

\section{Experiments and Results}
\label{s:pdf-exp}
As there are two classes (malware and benign) associated with the samples of our dataset, we conducted 10-fold cross-validation binary classification experiments to show the efficacy of our selected features. 
We considered various standard classifier models such as K-Nearest Neighbors (KNN), Random Forest (RF), Logistic Regression (LR), Linear Discriminant Analysis (LDA), Decision Tree (DT), Gaussian Naive-Bayes (GNB), Support Vector Machines (SVM) and XGBoost (XGB).
However, we report the classification accuracies for only KNN and RF models here as they were determined to be among the best performing models for various parametric settings and with the least variance across multiple random trials.
As the dataset has nearly a balanced distribution of classes, standard accuracy is used and reported to evaluate the performance of the classification models.

For binary analysis-based classification, we extract image, audio and hash  features described in Section~\ref{ss:pdf-exp1} for all the PDF samples in our dataset and perform classification using different classifiers. We tabulate the classification accuracy numbers in Table~\ref{t:bg} for KNN and RF models. We also provide the accuracy numbers for a voting ensemble classifier (VEC) that uses hard majority rule voting of the KNN and RF predictions. 
Overall, the image features performed better than that of the audio features and hash features.
\begin{table}[!htbp]
\scalebox{0.87}{%
\begin{tabular}{|c|c|c|c|c|}
\hline
\textbf{Feature} & \textbf{\# Dimensions} & \textbf{KNN} & \textbf{RF} & \textbf{VEC* (KNN+RF)} \\ \hline
Byteplot GIST~\cite{malware-images}    & 320                    & 0.9731     & 0.9671    & 0.9736               \\ \hline
Bigram-dct GIST~\cite{mohammedmalware}  & 320                    & 0.9794     & 0.9818    & 0.9821               \\ \hline
Melspectrogram~\cite{nataraj2020malware}   & 128                    & 0.9408     & 0.9694    & 0.9622               \\ \hline
MFCC~\cite{nataraj2020malware}             & 20                     & 0.9469     & 0.9463    & 0.9539               \\ \hline
Chroma~\cite{nataraj2020malware}           & 12                     & 0.8555     & 0.8893    & 0.8814               \\ \hline
Ssdeep~\cite{kornblum2006identifying}           & 40                     & 0.8138     & 0.9448    & 0.9213               \\ \hline
\end{tabular}%
}\\

{*VEC - Voting Ensemble Classifier}
\caption{Classification accuracies on different binary analysis-based (image, audio and hash) features.}
\label{t:bg}
\end{table}

For keyword analysis-based classification, we extract the structural features as described in Section~\ref{ss:pdf-exp2}.
We tabulate the binary classification results in Table ~\ref{t:bg2} below. 
The improved results over the binary analysis results suggest that the bag-of-words model features are better than byte-structure features, both in terms of feature complexity (25 dimensions vs. 320 dimensions) and classification accuracies (\texttt{99.51\%} vs. \texttt{98.21\%}).
\begin{table}[!htbp]
\scalebox{0.87}{%
\begin{tabular}{|c|c|c|c|c|}
\hline
\textbf{Feature} & \textbf{\# Dimensions} & \textbf{KNN} & \textbf{RF} & \textbf{VEC* (KNN+RF)} \\ \hline
Structural~\cite{Pareek2013MaliciousPD}    & 25                    & 0.9910     & 0.9947    & 0.9951               \\ \hline
\end{tabular}%
}\\

{*VEC - Voting Ensemble Classifier}
\caption{Classification accuracies on keyword analysis-based (risky tags) features.}
\label{t:bg2}
\end{table}

\begin{figure*}[!htbp]
    \centering
    \includegraphics[width=0.95\textwidth]{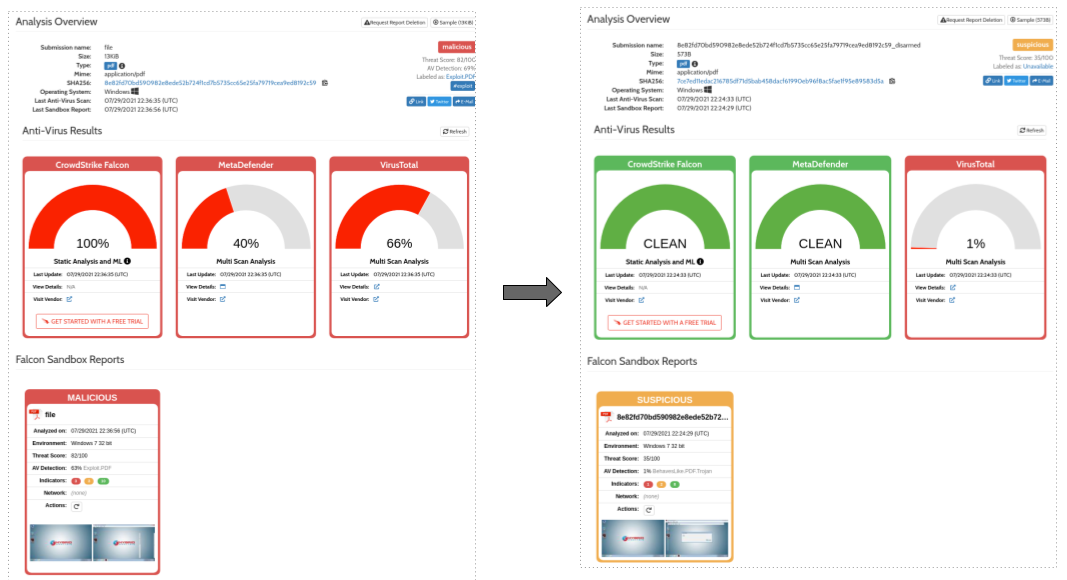}
    \caption{HybridAnalysis~\cite{hybridanalysis} overview reports of a PDF malware file (MD5 hash: \textit{	499bdcc2dab9ca4aaa49e40f1d1516d3}) and its obfuscation-removed counterpart (MD5 hash: \textit{	5a9ccbca186a7ceff12cfcabff4ab7a4}) --  the obfuscation removal was done using the 2$^{nd}$ method (Section~\ref{sec:pdf-deob}) and the reports indicate that the modified PDF malware was undetected by most AV solutions.}
    \label{fig:ha}
    \vspace*{-0.2in}
\end{figure*}
As described in Section~\ref{sec:pdf-combi}, the JFS metric is used to determine the best combination of features for holistic classification.
In our experiments, we find that the best combination of features include the image-based bigram-dct GIST~\cite{mohammedmalware}, the audio-based MFCC~\cite{nataraj2020malware} and the keyword-based structural~\cite{Pareek2013MaliciousPD} features with a JFS value of \texttt{0.87}. 
As there is a small room for performance improvement in the accuracy numbers, a simple feature concatenation is used as the proposed approach for combining features.
The concatenated (bigram-dct GIST + MFCC + structural) features produced an overall accuracy of \texttt{99.92\%} on the same PDF dataset. 
The other combinations performed slightly worse than this combination, which is reflective of the orthogonality measure of the features from the JFS metric.

Inspired from the bag-of-words model from keyword analysis, we proposed a dynamic analysis-based malware detection method as described in Section~\ref{sec:dynamic} to get the API call sequence features for PDF files. 
We tabulate the binary classification results for different classification models using these features in Table~\ref{t:bg3} below.
Counter-intuitively, the dynamic method performed slightly worse than our static-based holistic approach in terms of classification accuracies (\texttt{99.18\%} vs. \texttt{99.92\%}) which also shows the efficacy of the latter. 
If dynamic methods are of interest, then the proposed approach demonstrates its usefulness for PDF malware detection, and has promising directions for future research. 
\begin{table}[!htbp]
\scalebox{0.87}{%
\begin{tabular}{|c|c|c|c|c|}
\hline
\textbf{Feature} & \textbf{\# Dimensions} & \textbf{KNN} & \textbf{RF} & \textbf{VEC* (KNN+RF)} \\ \hline
API call sequences   & 257                    & 0.9881    & 0.9901    & 0.9918               \\ \hline
\end{tabular}%
}\\

{*VEC - Voting Ensemble Classifier}
\caption{Classification accuracies on dynamic analysis-based (API calls) features.}
\label{t:bg3}
\end{table}

For adversarial analysis, we upload the modified PDF files generated from the obfuscation removal methods described in Section~\ref{sec:pdf-deob} to Virustotal~\cite{virustotal} and HybridAnalysis~\cite{hybridanalysis} so as to get the malware reports and observe the trends.
In most cases, the obfuscation-removed PDF samples were able to fool most Antiviruses and left undetected as shown in Figure~\ref{fig:ha}.   
The reports also suggest that (1) some of the antivirus engines (list includes Kaspersky, ESET-NOD32, ZoneAlarm, etc.) look for risky object tags irrespective of the tag’s case type, and (2) some of the engines (list includes Baidu, Sophos, McAfee, etc.) look for presence of malicious code irrespective of whether it is executed or not.
Our holistic approach (Section~\ref{sec:pdf-combi}) classifies \emph{all} the modified PDF malware samples as malware (and the same with benign samples), which helps users to prevent the risk of running the hidden malicious code within such PDF files.
This is expected as the concatenated model has pre-dominantly binary features that are derived from the whole PDF file which are hardly perturbed with small delta modifications made to the PDF code.

\section{Conclusion}
\label{s:conc}
In this paper, we present and practically demonstrate a novel holistic approach to PDF malware detection. 
We use image, audio and hash representations of PDF binaries for signal-based malware analysis and a bag-of-words model for PDF structural analysis. 
Additionally, we also develop a bag-of-words model using the API call sequences from the dynamic analysis reports for PDF malware detection.
We then use obfuscation removal techniques to augment the PDF malware dataset and to demonstrate how most of the antiviruses fail to detect such adversarially generated samples. 
Simple combination of various features demonstrate orthogonality among the different feature space models. 
We strongly believe that this holistic approach lays the groundwork of a fundamental building block for threat intelligence platforms that aim at protecting systems from diverse attacks. 
Though we conduct experiments only with the PDF files, we expect that the approach can be applicable to other types of data that contain byte-streams and file structure. 
Therefore, as a future work, we will collect data of other malicious document types (e.g., .rtf, .docx, .xlsx files) and perform further investigation.

\input{main.bbl}

\end{document}